\documentclass[runningheads]{svmult}

\usepackage{makeidx}   % allows index generation
\usepackage{graphicx}  % standard LaTeX graphics tool
\usepackage{subeqnar}  % subnumbers individual equations
\usepackage{multicol}  % used for the two-column index
\usepackage{physprbb}  % modified textarea for proceedings,
\makeindex             % used for the subject index
\begin{document}
\title*{Dynamical models linking BH masses and DM content}
\toctitle{Linking the black hole and bulge formation}
% allows explicit linebreak for the table of content
%
%
\titlerunning{Linking the black hole and bulge formation}
% allows abbreviation of title, if the full title is too long
% to fit in the running head
%
\author{Pieter Buyle\inst{}
\and Herwig Dejonghe\inst{}
\and Maarten Baes\inst{}
}
\authorrunning{Pieter Buyle et al.}
% if there are more than two authors,
% please abbreviate author list for running head
%
%
\institute{Ghent University, Krijgslaan 281 S9, Ghent B-9000, Belgium}

\maketitle              % typesets the title of the contribution

\begin{abstract}
  We investigate the relation between the dark matter distribution of
  galaxies and their central supermassive black holes which is
  suggested by the $v_c-\sigma$ relation. Since early-type galaxies
  appear to have larger black holes than late-type ones, we look for
  an equivalent pattern in the dark matter distribution as a function of
  Hubble type. To achieve our goal we use a state-of-the-art modelling
  code that allows a variety of geometries to be fitted to a
  combination of radio and optical observations of galaxies with
  different morphology.
\end{abstract}

\section{Introduction}
Current models for galaxy formation and evolution all incorporate the
role of a supermassive black hole. More in particular, they are
usually calibrated against known observational relations, such as the
Tully-Fisher, $M_{BH}-\sigma$ [3][5] and the $v_c-\sigma$ [1][4]
correlations. Since it becomes clear that the $v_c-\sigma$
relation is gaining importance due to its putative redshift independence, a
firm observational basis for it seems to be called for.  However, at
the moment the relation is based on only 28 galaxies making it
impossible to measure the intrinsic scatter. Moreover, its precise
meaning is not established at all, let alone the combination of it
with the $M_{BH}-\sigma$ relation. This in turn points to a poorly
understood connection between the dark matter distribution and the SBH
(Fig. 1):
\begin{equation}
\log{\left(\frac{M_{BH}}{M_\odot}\right )}=(4.21\pm 0.60)\log{\left (\frac{v_c}{u_0}\right )}+(7.24\pm 0.17),\ \ \ \ u_0=200\ km\ s^{-1}.
\end{equation}
\section{Intrinsic scatter of the $v_c$-$\sigma$ relation}
We started a project to measure the central velocity dispersions of 20 spiral galaxies with the 3.5m telescope at
Calar Alto and we recently observed 12 Low Surface Brightness (LSB) galaxies with the VLT [2]. The majority of the LSB galaxies of our sample have asymptotic circular velocities below 100\ km\ s$^{-1}$, while the velocities of the 20 spiral galaxies of our second sample are all between 100\ km\ s$^{-1}$ and 150\ km\ s$^{-1}$. In combination with our already obtained data with velocities above 150\ km\ s$^{-1}$ [1], this will allow us to obtain the real intrinsic scatter of the $v_c-\sigma$ relation and to investigate this relation in the low mass regime, where it is hinted by small-number statistics to deviate from the known relation. The sample has also a nice spread along the Hubble sequence and therefore it allows us to investigate any morphological dependence as well. 
\begin{figure}[b]
\begin{center}
\includegraphics[width=.4\textwidth]{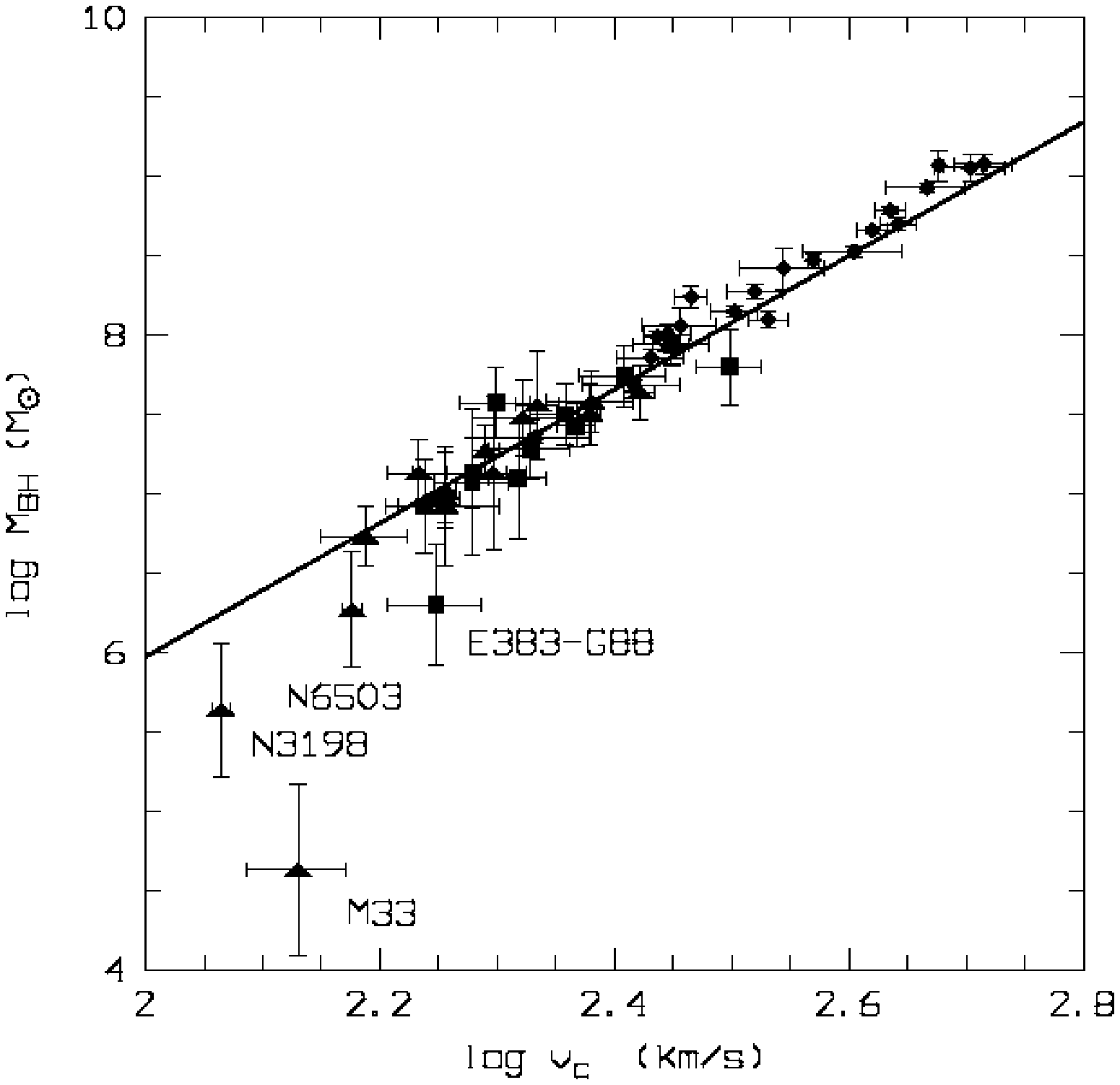}\hspace{1cm}
\includegraphics[height=4.5cm,width=.4\textwidth]{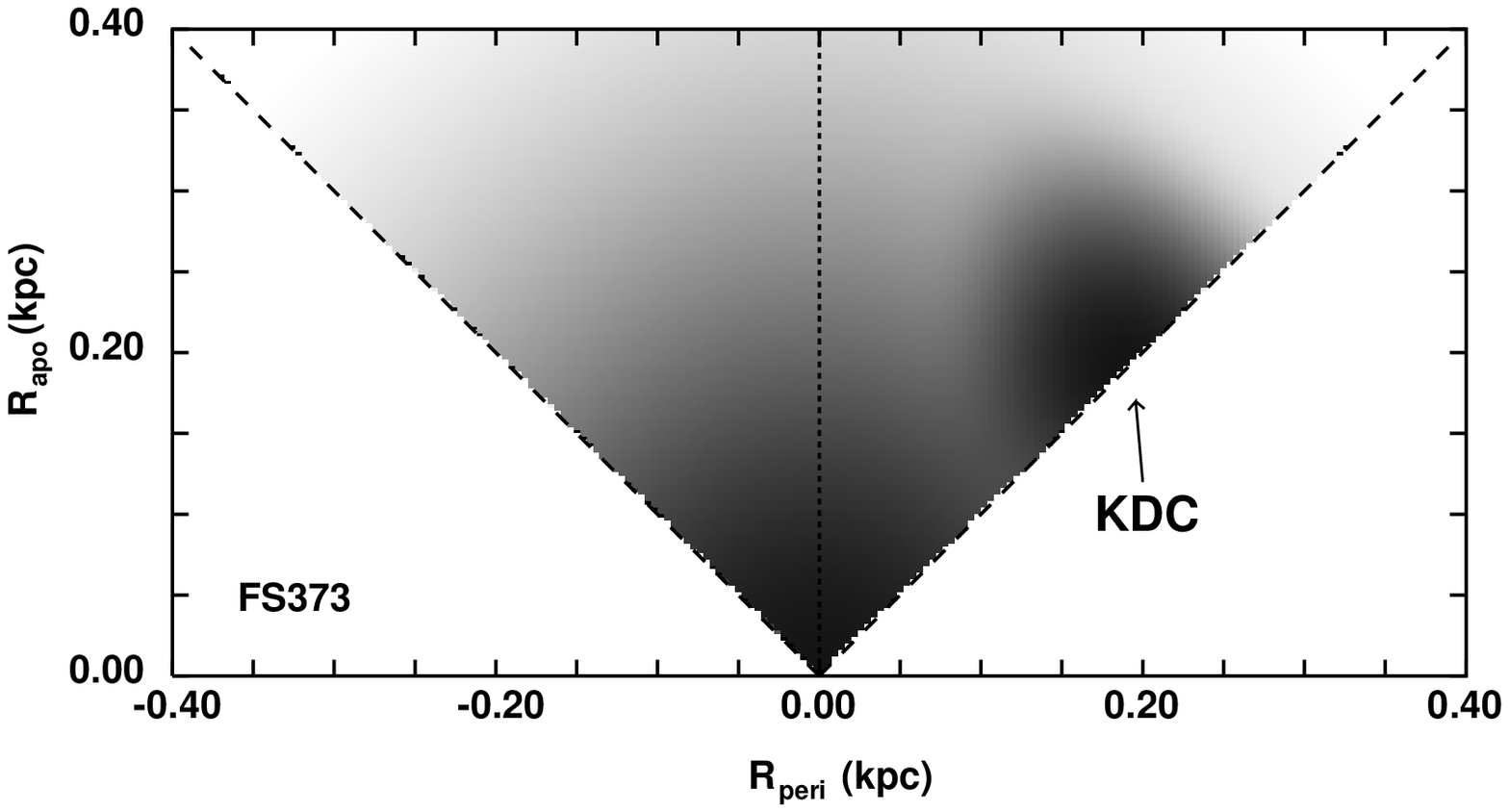}
\end{center}
\caption[]{{\bf (a)} The $v_c-M_{BH}$ relation (1). Ferrarese's data (2002) are represented by triangles, Kronawitter's (2000) by circles and Baes et al.'s by squares. {\bf (b)} Optical based distribution function of FS373 (De Rijcke et al., 2004, accepted for A\&A)}
\end{figure}
\section{Dark matter distribution}
Almost all information that we have about the dark matter distribution
within galaxies is based on rotation curves only and models that
assume a spherical dark matter distribution. We plan to improve on the modelling part with a code which allows for both spherical and
axisymmetric geometries to be fitted to a combination of radio and
optical observations. The aim is to fit a model
directly to the radio observations including the full emission
distribution along the line-of-sight at each point without making preliminary assumptions of the
circular velocity. Models will be fitted to a
sample of 12 Low Surface Brightness galaxies of which HI observations
are being made at the ATCA and of which we will propose photometry at
the AAT to be able to derive the contribution of the luminous and dark
matter. Instead of measuring rotation curves, we will derive
distribution functions which contain all kinematical information of
the system. In particular, we plan to investigate the link between the
distribution function and morphology in our Low Surface Brightness
sample, with the aid of diagrams in turning point space, such as Fig.2.

%INDEX%%%%%%%%%%%%%%%%%%%%%%%%%%%%%%%%%%%%%%%%%%%%%%%%%%%%%%%%%%%%%%%
% Please check with the editor of your book whether he plans to
% include a "mutual" subject index - if so, please code your entries
% in the standard syntax. For your own purposes you may print your
% "personal" index by using the following commands:
%
%\clearpage
%\addcontentsline{toc}{section}{Index}
%\flushbottom
%\printindex
%%%%%%%%%%%%%%%%%%%%%%%%%%%%%%%%%%%%%%%%%%%%%%%%%%%%%%%%%%%%%%%%%%%%%

\end{document}